\documentclass{actasen}

\begin{document}

%%ÉèÖÃÊ×Ò³Ò³Âë
\setcounter{page}{115}

\Volume{2015}{35}% Äê¡¢¾í

%%ҳüÉèÖÃ

\runheading{LI Zhao-Sheng}%

\title{Constraints on the mass and radius of neutron stars from X-ray observations$^{\dag}~ \!^{\star}$}

\footnotetext{$^{\dag}$  Supported by China Postdoctoral Science Foundation (2014M560844).

Received ; revised version 

$^{\star}$ A translation of {\it Acta Astron. Sin.~}
Vol. 55, No. 2, pp. 116--126, 2014 \\
\hspace*{5mm}$^{\bigtriangleup}$ lizhaosheng@pku.edu.cn\\

\noindent 0275-1062/01/\$-see front matter $\copyright$ 2011 Elsevier
Science B. V. All rights reserved. %%

\noindent PII: }

\enauthor{LI Zhao-Sheng$^{\bigtriangleup}$ }{Department of Astronomy, Peking University}

\abstract{This article gives a very brief introduction about measuring the mass and radius of neutron star from X-ray observations. The masses and radii of neutron stars can be determined from photospheric radius expansion  bursts in low-mass X-ray binaries, X-ray pulse profile modeling in accreting X-ray pulsars, gravitational redshift measurement in low-mass X-ray binaries and thermal X-ray spectral fitting in quiescent low-mass X-ray binaries. }


\maketitle

\section{Introduction}
The equation of state (EoS) of superdense matter is quite uncertain in astrophysics and nuclear physics. Neutron stars (NSs), one of the most condense stars in the Universe, provide us an unique opportunity to approach it. Generally, two categories of EoS were proposed, which can produce gravity-bound NSs and self-bound quark stars or quark-cluster stars, respectively.  The measurements of the radius and mass of NS can provide useful information to test various theoretical EoSs. Dozens of NSs have accurate mass measurement in binary systems, such as double NSs system or white dwarf--neutron star system.  The precise measurement of the radius of NS, however, is far beyond achievement. The measurement of NS radius is very critical for EoS constraining. Especially, Fortin et al. (2014)  \cite{fortin14} claimed that the NSs with mass in the range  $1.0-1.6~M_{\odot}$ should be larger than 12 km; otherwise, the presence of hyperons in NS cores predicted by the relativistic mean field approximation is ruled out. And then, the so-called hyperon puzzle arises (e.g., Bednarek et al. 2012 \cite{bednarek12}). 

\section{Methods}
Several methods to measure the mass and radius of NS are introduced as follows:

(1). The most straightforward way of determining the mass and radius ratio is by measuring the gravitational redshift of spectral lines produced on the neutron star surface. This observation puts very tight constraint on the mass-to-radius ratio, since, the redshift can be expressed as $z=1-(1-2GM/Rc^2)^{-1/2}$, where, $M$ and $R$ are the mass and radius of a NS respectively, $G$ is gravitational constant, $c$ is the speed of light. Cottam et al. (2002) \cite{cottam02} reported the detection of gravitation redshift with $z=0.35$ in EXO 0748-676 during its X-ray burst epochs. Unfortunately, this gravitational redshift wasn't confirmed by the following observations\cite{cottam08}. So, it needs to be checked by further powerful X-ray telescopes.
 
(2). In accreting induced X-ray pulsars, X-ray pulses are modulated by general relativistic light-bending effect, which depends on the compactness $M/R$, special relativistic Doppler boosting and aberration. Modeling the X-ray pulse profile, the mass and radius can be constrained. Here, lots of processes should be considered carefully, such as geometrical factors and NS atmosphere emission\cite{psaltis14}.

(3). Almost of all LMXBs are transients in X-ray band. In accretion phase, LMXBs behavior bright X-ray emission, most of energy emit outwards and small parts of energy deposit in the inner crust of NS. When accretion terminated, the LMXB transit into quiescent state. The hotter inner crust radiate outwards as well as inwards. And then, its X-ray spectrum is dominated by thermal X-ray emission, which is well fitted by nsatmo in \textit{Xspec} package. With the known distance to the source, e.g. in globular cluster, the mass and radius are easy to obtained from high signal-to-noise X-ray spectra\cite{guillot13}. 

(4). Type I X-ray bursts, which is also known as Thermonuclear X-ray bursts, in LMXBs are a sudden energy release process, which lasts tens to hundreds of seconds and can emit as high as Eddington luminosity ($\sim3.79 \times 10^{38} ~\rm{erg/ s}$ ). In the classical view, type I X-ray bursts are powered by the unstable thermonuclear burning of H/He accreted on the NS surface through its companion star Roche lobe overflowing. Most of the spectra of type I X-ray bursts can be well fitted by a pure blackbody spectrum. Photospheric radius expansion (PRE) bursts, a special case of type I X-ray bursts, were phenomenally distinguished from the time-resolved spectra. At the touchdown moment, where the blackbody temperature and its normalization reach their local maximum and minimum during X-ray burst, respectively, the referred bolometric luminosity corresponds to its Eddington luminosity, that is, the radiation pressure is balanced by gravity. After the touchdown point, the residual thermal energy is believed to cool on the whole surface of the NS during the burst tail. So, the mass and radius of the NS could be constrained if the distance to the source was measured independently, i.e., also in globular clusters \cite{li15}.

\section{The mass and radius of NS in 4U 1746-37}

4U 1746-37 is a low mass X-ray binary (LMXB) located in the Globular Cluster NGC 6441. The distance to NGC 6441 is $11.0^{+0.9}_{-0.8} ~\mathrm{kpc}$. From the type I X-ray burst catalog of RXTE, three PRE bursts were identified 4U 1746-37. We simulated its mass and radius as shown in Figure 1, which is an ultra-low-mass quark-cluster star candidate. Please see the details in Li et al. (2015)\cite{li15}.

\begin{figure}[tbph]
\begin{minipage}{0.6\textwidth}
\includegraphics[bb=50 315 525 595,width=7.0cm]{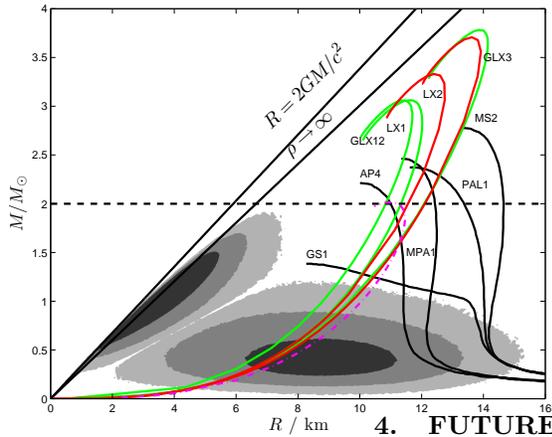}
\end{minipage}%
\begin{minipage}{0.4\textwidth}
\caption{The 1-, 2-, 3-sigma $M-R$ confidence regions of 4U 1746-37.
The dashed line denotes two observed near $2M_\odot$ NSs. The left black lines show the the general relatively (GR) limit and the central density limit, respectively. Theoretical mass-radius relations for several NS EoS models are displayed, colored curves for self-bound quark-cluster stars (dashed line for quark star) and black curves for gravity bound NSs}
\end{minipage}
\end{figure}

\section{Future Works}

X-ray Timing and Polarization (XTP) mission, the next generation X-ray telescope, has a capability of measure the mass and radius with much smaller uncertainties from above mentioned methods.

\acknowledgements{Z. S. Li is supported by China Postdoctoral Science Foundation (2014M560844). }

\end{document}